\documentclass[preprint,12pt]{elsarticle}

\usepackage{graphicx}
\usepackage{float}
\usepackage{amssymb}
\usepackage{amsmath}

\usepackage{multicol}
\usepackage{multirow}
\usepackage{tabularx}
\usepackage{color,soul}

\usepackage{comment}
\usepackage{natbib}
\usepackage{mathtools}
\usepackage{caption}
\usepackage{subcaption}
\usepackage[colorinlistoftodos]{todonotes}
\usepackage[colorlinks=true, allcolors=blue]{hyperref}

\usepackage{lineno}

\sethlcolor{yellow}

\journal{Building Simulation}

\begin{document}
\begin{frontmatter}

\title{Modeling ventilation in a low-income house in Dhaka, Bangladesh}


\author[stanford_welab]{Yunjae Hwang \corref{label1}}
\author[berkeley]{Laura(Layla) H. Kwong}
\author[icddrb]{Mohammad Saeed Munim}
\author[icddrb]{Fosiul Alam Nizame}
\author[stanford_medicine]{Stephen P Luby}
\author[stanford_welab]{Catherine Gorl\'e}
\address[stanford_welab]{Department of Civil and Environmental Engineering, Stanford University, CA, USA}
\address[berkeley]{School of Public Health, University of California, Berkeley, CA, USA} 
\address[icddrb]{International Centre for Diarrhoeal Disease Research, Bangladesh (icddr,b), 68 Shaheed Tajuddin Ahmed Sarani, Mohakhali, Dhaka, Bangladesh}
\address[stanford_medicine]{School of Medicine, Stanford University, CA, USA \vspace{-20pt}}
\cortext[label1]{Corresponding author, yunjaeh@stanford.edu}

\begin{abstract}
According to UNICEF, pneumonia is the leading cause of death in children under 5. 70\% of worldwide pneumonia deaths occur in only 15 countries, including Bangladesh. Previous research has indicated a potential association between the incidence of pneumonia and the presence of cross-ventilation in slum housing in Dhaka, Bangladesh. The objective of this research is to establish a validated computational framework that can predict ventilation rates in slum homes to support further studies investigating this correlation. To achieve this objective we employ a building thermal model (BTM) in combination with uncertainty quantification (UQ). The BTM solves for the time-evolution of volume-averaged temperatures in a typical home, considering different ventilation configurations. The UQ method propagates uncertainty in model parameters, weather inputs, and physics models to predict mean values and 95\% confidence intervals for the quantities of interest, namely temperatures and ventilation rates in terms of air changes per hour (ACH). The model predictions are compared to on-site field measurements of air and thermal mass temperatures, and of ACH. The results indicate that the use of standard cross- or single-sided ventilation models limits the accuracy of the ACH predictions; in contrast, a model based on a similarity relationship informed by the available ACH measurements can produce more accurate predictions with confidence intervals that encompass the measurements for 12 of the 17 available data points. 
\end{abstract}

\end{frontmatter}

\section{Introduction}
Pneumonia is the leading cause of death in children under five \citep{gbd2016}, accounting for 18\% of total mortality in this age group~\citep{qazi2015}. Worldwide, approximately one million children die from pneumonia every year, with 70\% of cases occurring in only 15 countries, one of which is Bangladesh \citep{unicef2016}.
In Dhaka, the capital of Bangladesh, 4 million people are estimated to live in slums or informal settlements. These communities have poor quality housing \citep{un2010state} and are densely crowded: more than 95\% of homes are one room dwellings with a surface area of less than 14 square metres \citep{islam2006slums}. The high population density combined with inadequate healthcare causes serious health risks for slum residents \citep{ahmed2014factors}. In 2014, Ram, et al.~\citep{ram2014household} reported an association between the incidence of pneumonia and ventilation status of Dhaka homes: households where pneumonia occurred were 28\% less likely to be cross-ventilated. It is conceivable that improved cross ventilation could dilute respiratory pathogens and indoor air pollution. On the other hand, the ambient air in Dhaka is highly polluted, so cross ventilation may decrease air quality~\citep{rahman2019recent, siddiqui2020chronic}. A rigorous exploration of the relationship between ventilation and pneumonia requires an accurate estimation of a household's ventilation rate.

A range of computational tools with widely varying computational cost and levels of fidelity in the representation of the physics can be used to model natural ventilation. On the high-cost, high-fidelity end, computational fluid dynamics (CFD) models can solve for the three-dimensional flow and temperature field in a specific geometry with well-defined boundary conditions. CFD has been used extensively for natural ventilation modeling~\citep{jiang2001study,evola2006computational,hu2008cfd,ramponi2012cfd,lo2013effect, gilani2016cfd,lamberti2018uncertainty,liu2019cfd, shirzadi2020cfd}, but the model complexity and computational cost introduce limitations when a variety of configurations under highly variable boundary conditions have to be analyzed~\citep{etheridge2015}. 
On the lower-cost, lower-fidelity end, building simulation tools such as EnergyPlus couple a building thermal model (BTM) with a zonal airflow model to predict indoor temperatures and ventilation rates \citep{crawley2001energyplus,schulze2013controlled,zhai2011assessment,taheri2016performance,martins2016validation}. The reduction in both cost and fidelity is primarily related to the use of reduced-order airflow models, which employ empirical relations to define flow rates between different zones. In their simplest and lowest cost form, these models employ a single zone, such that the ventilation rates are determined by an envelope flow model, i.e. a model that quantifies the net airflow rate through the ventilation openings. The primary advantage of this type of model is its very low computational cost, making it ideal for analyses that require repeated simulations, such as the evaluation of multiple ventilation configurations with uncertainty quantification (UQ). A modeling framework using a BTM with UQ has been shown to accurately predict indoor air temperatures in a three-story office building with buoyancy-driven natural ventilation~\citep{lamberti2018uncertainty}, but validation of ACH predictions in configurations with combined wind- and buoyancy-driven flow remains to be investigated.

The objective of this research is to validate a low-cost modeling framework for predicting indoor temperatures and ventilation rates in slum homes, such that the resulting framework can support further investigation of the association between pneumonia and household ventilation. The framework encompasses two major components: (1) a building thermal model (BTM) that solves for the volume-averaged temperature in a slum residence and calculates the ventilation rate in terms of air changes per hour (ACH), and (2) a UQ method that propagates the effect of uncertainties due to model parameters, weather conditions, and physics models to provide predictions with quantified confidence intervals (CI). The resulting framework will predict the mean and 95\% CI of two primary quantities of interest (QoI), i.e. the indoor temperature and ACH. 

The validation test case consists of a representative single-room house rented in an urban slum of Dhaka, Bangladesh. Various opening types and sizes are considered: a skylight, a floor-level vent, a ceiling-level vent, and a large window. Four different combinations of these openings are tested to investigate the resulting differences in measured ventilation rates and model accuracy.
Field measurements were obtained during a 15 day period. Indoor and outdoor temperatures, as well as wind speed and direction at roof height and at free-stream level, were obtained. In addition, 19 tracer concentration decay measurements were performed to measure the ACH at different points in time for the four ventilation configurations. Model validation is performed by comparing indoor temperature and ACH predictions to the field measurements. In addition, the analysis of the results includes a sensitivity analysis to identify the relative importance of the different uncertain parameters and identify future research efforts to reduce model uncertainty.

In the remainder of this paper, Section 2 first presents the configuration of the urban slum home and the experimental setup for the temperature, wind speed and direction, and ventilation rate measurements. Subsequently, Section 3 presents the computational framework with the building thermal model and uncertainty propagation methods. Lastly, Section 4 presents the validation of the computational framework by comparing the results to the field measurements and the sensitivity analysis. Finally, the conclusions are presented in Section 5.

\section{Test case and field measurements} \label{field experiment}
In this section, we first present the test house geometry and the different ventilation configurations. Subsequently, we discuss the field measurements performed during a 15 day period in February 2019. Outdoor temperature and wind measurements were performed to characterize the model input parameters, while indoor temperature and ACH measurements were performed to support model validation. 

\subsection{Configuration of the test house} \label{house configuration}
\begin{figure}[htbp]
\centering
\begin{subfigure}[t]{0.5\linewidth}
    \centering
    \includegraphics[width=0.9\linewidth]{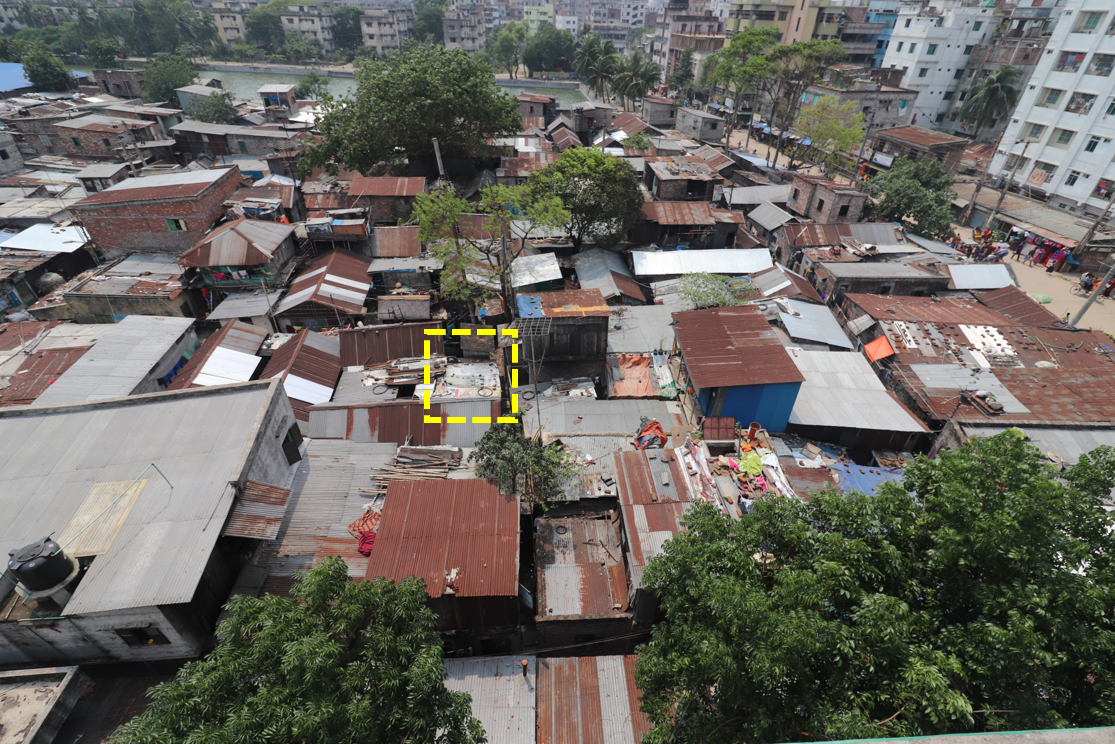}
    \caption{ }
    \label{fig:housepicture}
\end{subfigure}%
~
\begin{subfigure}[t]{0.49\linewidth}
    \centering
    \includegraphics[width=0.9\linewidth]{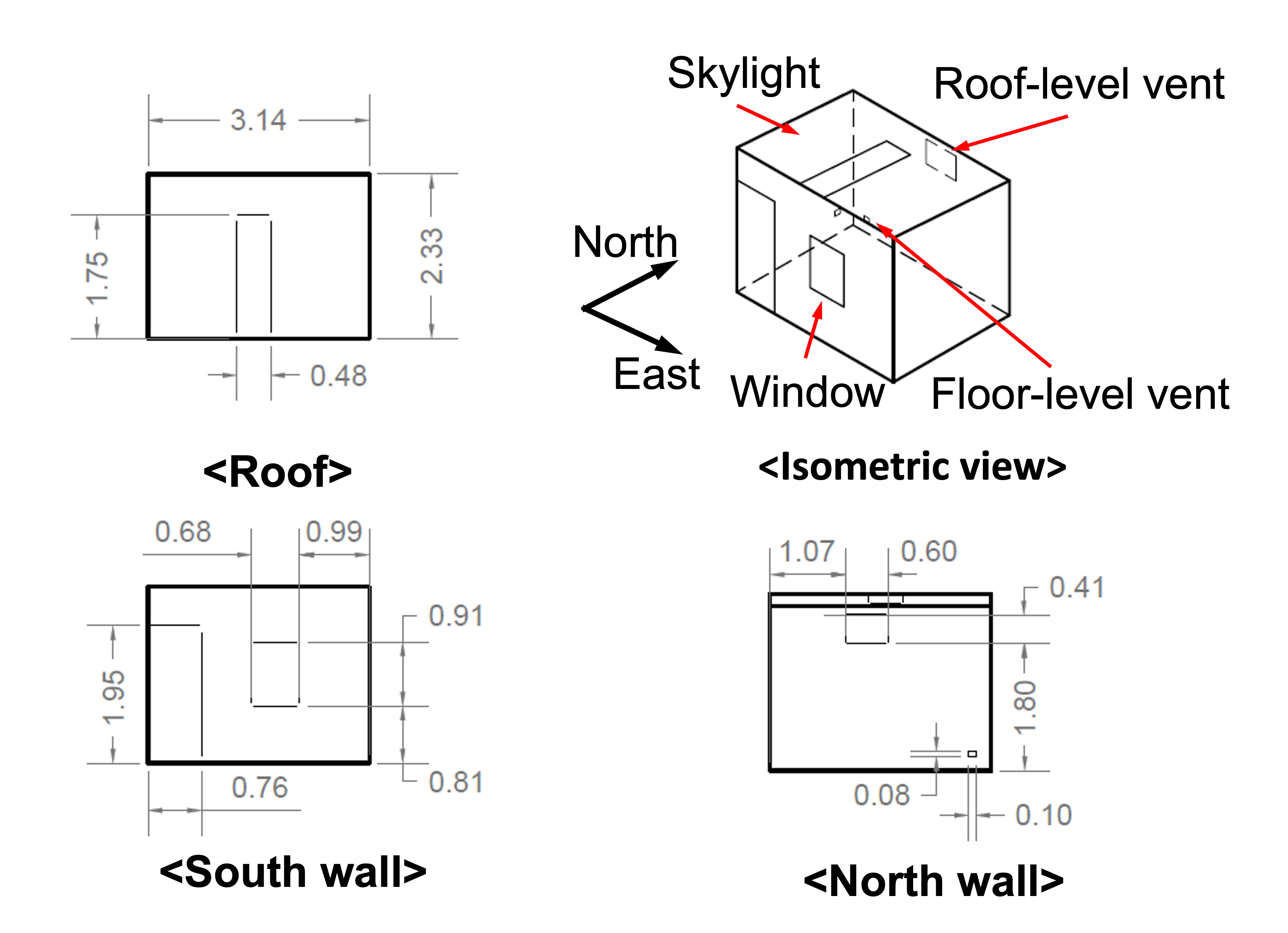}
    \caption{ 
    }
    \label{fig:house_configuration}
\end{subfigure}
\caption{Study site of the current study: (a) a bird-eye view of the Outfall urban-slum area in Bangladesh; (b) drawings of the target house with the four openings for testing ventilation configurations}
\end{figure}

The measurements were performed in a rental house in the Outfall slum of Dhaka, shown in Figure~\ref{fig:housepicture}. The home, constructed with brick walls and a tin roof (semi-pucca), had a rectangular floor plan of 3.14 m $\times$ 2.34 m and a slanted ceiling with a height of 2.51 m at the north wall and 2.33 m at the south wall. Figure~\ref{fig:house_configuration} depicts a sketch of the home, indicating the different ventilation openings constructed for the experiments. The south wall faces the neighborhood's public kitchen, with a door and a window of 0.68 m $\times$ 0.91 m with a security grill. The north wall has two vents: one larger roof-level vent measures 0.60 m $\times$ 0.41 m and is covered with a concrete frame and one smaller floor-level vent measures 0.09 m $\times$ 0.11 m. Lastly, an operable, tin-plate skylight, measuring 0.48 m $\times$ 1.75 m, was constructed in the roof. 
The experiments considered four configurations to test different combinations of two openings: (1) the skylight and floor-level vent open, (2) the skylight and roof-level vent open, (3) the window and roof-level vent open, and (4) the skylight and window open.  

\subsection{Outdoor temperature and wind measurements}
\begin{figure}[h!]
    \centering
    \includegraphics[width=\textwidth]{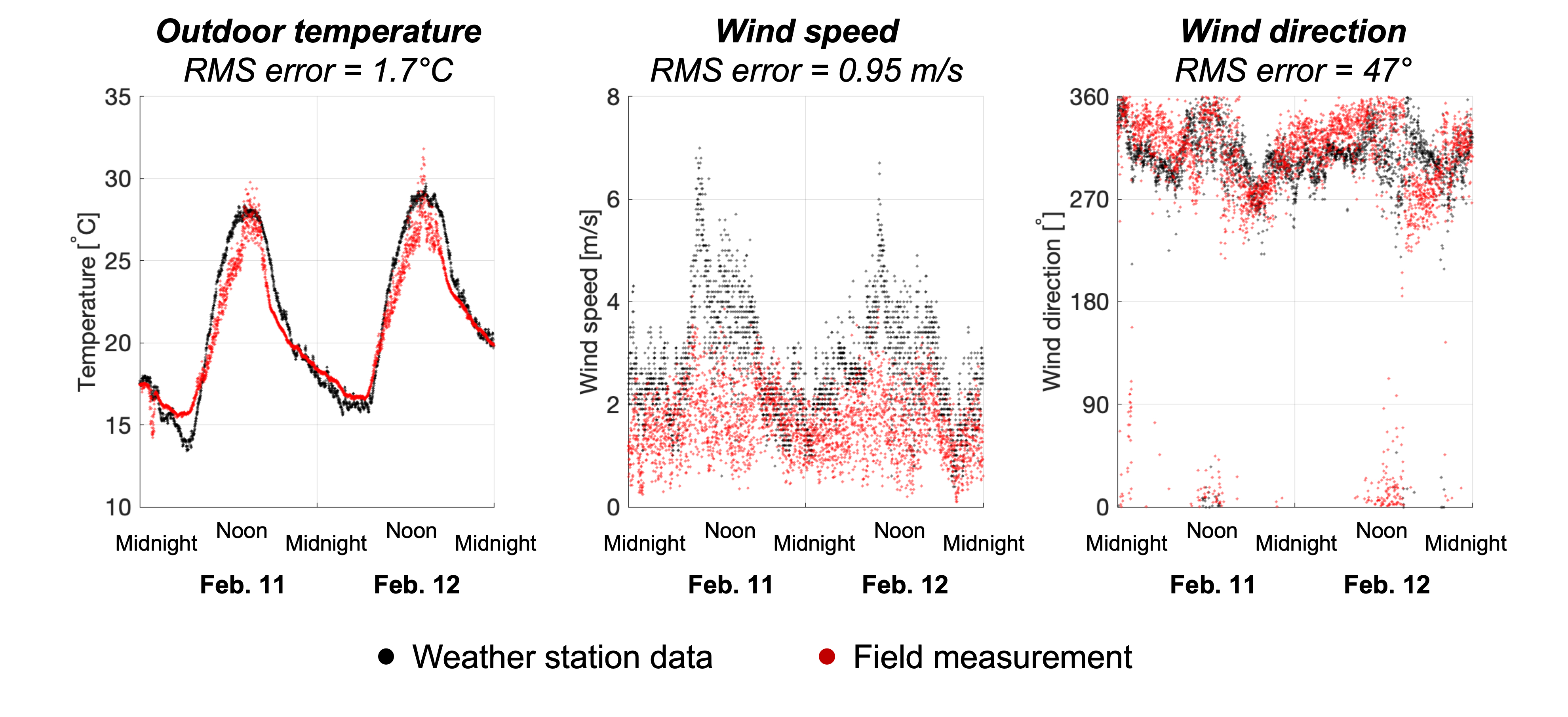}
    \caption{Verification of measurement: outdoor air temperature (left), wind speed (center), and wind direction (right)}
    \label{fig:measurement_verification}
\end{figure}

Natural ventilation flow is driven by indoor to outdoor temperature differences and wind conditions. As such, measurements of the outdoor temperature and wind speed and direction are needed to specify the BTM model inputs. A standard approach is to use data from a nearby weather station. For the target site, the closest station publishing meteorological data online each minute is located 14 km away, at Hazrat Shahjalal International Airport.
To ensure that the temperature and wind speed data recorded at the airport are representative of the local conditions, a mobile weather station was installed on the roof of the Maniknagaar Model School. At 25m high, this building is the tallest building in the surrounding area and it is located approximately 25 m away from the target house. The rooftop station records the outdoor temperature and wind velocities and directions with a sampling frequency of 1 Hz. Figure~\ref{fig:measurement_verification} shows the comparison of the measurements in both locations. Overall, the local rooftop measurement agrees well with the weather station data, capturing the varying trend of temperature cycles, the peaks of wind speed and the dominant wind directions, where the RMS error for the temperature, wind speed and wind direction between the two data are 1.7$^\circ$C, 0.95 m/s, and 47$^\circ$, respectively. Small discrepancy between the two data stems from the difference in the conditions of each measurement: our local measurement recorded the data at the center of a city and 14 km away from the weather station, which logs the weather station in open terrain. Based on this comparison, the model inputs for the wind speed and direction will be defined based on the local roof-top measurement, while the outdoor temperature and solar radiation data will be obtained from the airport weather station. The method used to post-process this data to define the input parameters for the UQ analysis with the BTM is presented in section~\ref{sec:input_uncertainty}. 

\subsection{Indoor air and thermal mass temperature measurements} \label{sec:temperature wind measurement}
\begin{figure}[htbp]
    \centering
    \includegraphics[width=0.8\textwidth]{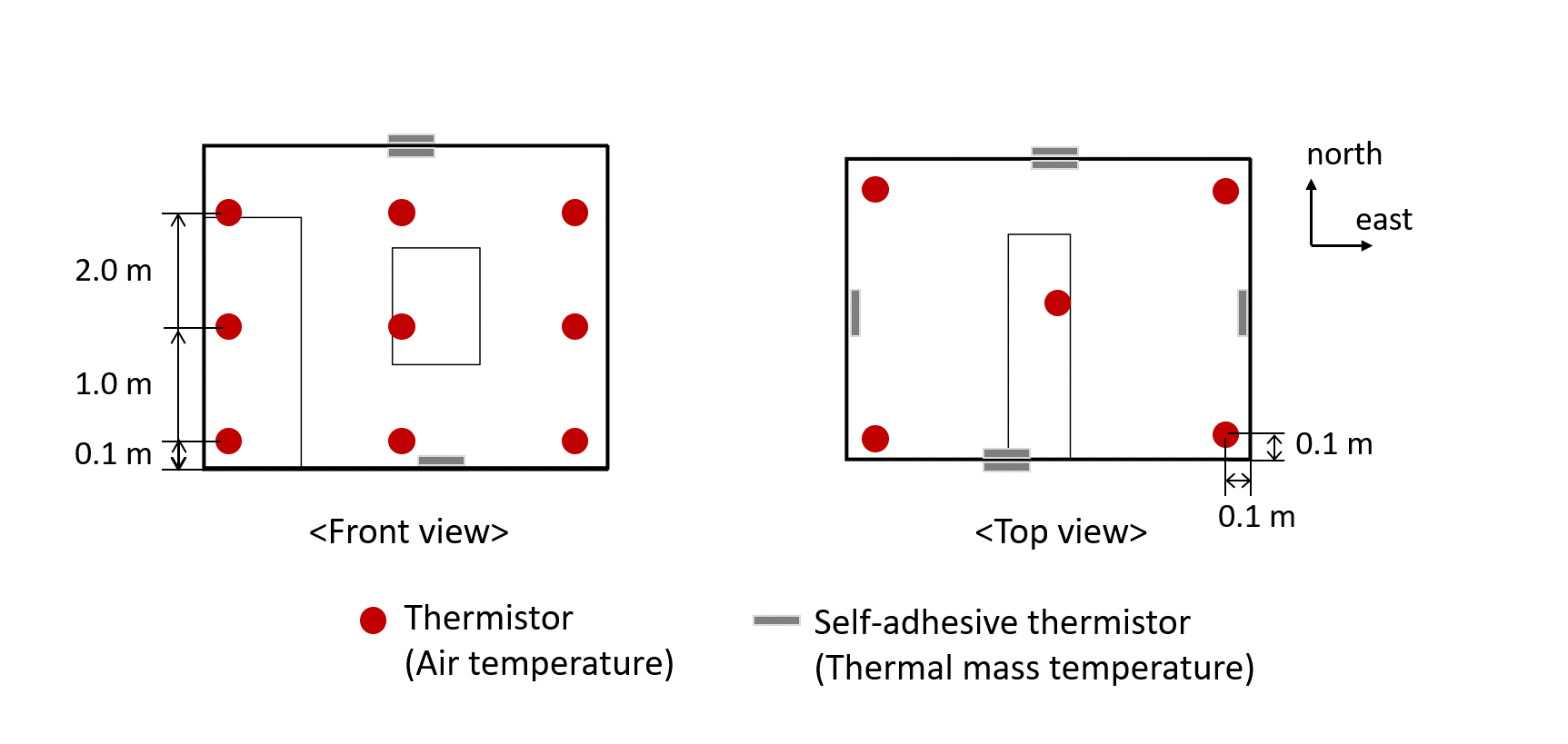}
    \caption{Sensor setup at the target slum house: thermistors for indoor air temperatures, self-adhesive thermistors for surface temperatures of thermal masses, and sonic anemometer for wind speed and direction at roof height}
    \label{fig:sensor_setup}
\end{figure}

Figure \ref{fig:sensor_setup} presents an overview of the sensor setup at the slum house, which includes 24 temperature sensors and one anemometer. The temperature sensors measure the indoor air temperature at 15 locations: 5 horizontal locations (4 corners and the center) at 3 different heights (0.1 m, 1.0 m and 2.0 m). In addition, thermal mass surface temperatures are measured using self-adhesive sensors at 9 locations: both the inner and outer surfaces of the roof, the north wall, and the south wall, and the inner surfaces of the floor, the east wall and the west wall. An anemometer was installed on top of the roof for assessing the local wind speed and direction, which are influenced by surrounding built environment.

The temperature and wind measurements are recorded using a data logger (CR300, Campbell Scientific, Inc.) with a sampling frequency of 1Hz. Wind speed data is obtained using a 2D anemometer (WindSonic, Gill Instruments, Ltd), which has an accuracy of 2\% at 12 m/s for both wind speed and direction. For the temperature measurements we use thermistors for two reasons: thermistors have an accuracy of $\pm$0.1$^\circ$C in the target range of 10$^\circ$C to 50$^\circ$C and they have limited sensitivity to the power outages that occur in the slum. All thermistors were calibrated with a dry block calibrator (ThermCal130, Accurate Thermal Systems LLC) to prevent self-heating from affecting the measurements. 

\label{temperature_measurement}
\begin{figure}[htb]
\centering
\includegraphics[width=0.8\linewidth]{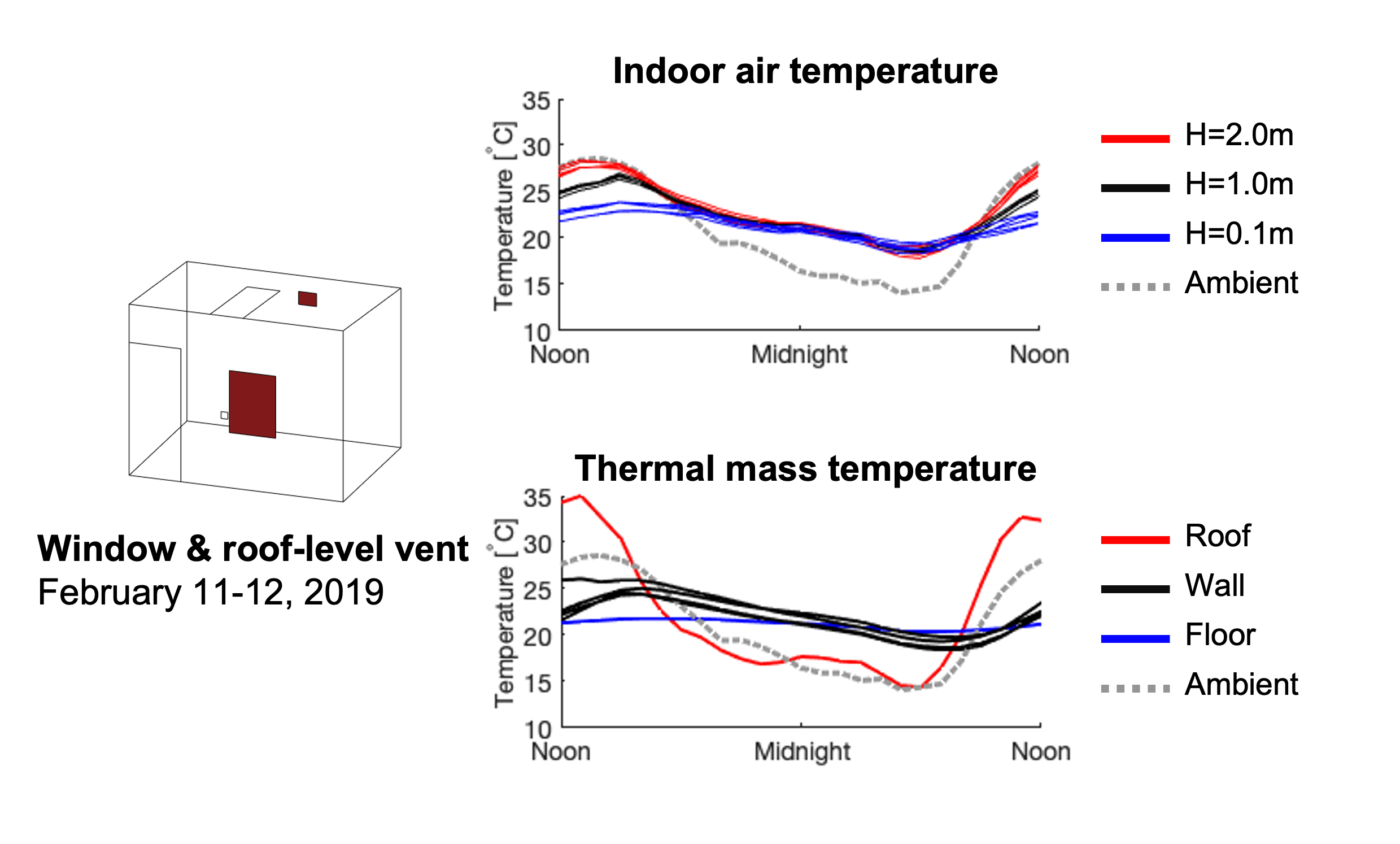}
\caption{Time series of indoor air temperatures measured at three different heights}
\label{fig:indoor_temperature}

\end{figure}
Figure \ref{fig:indoor_temperature} displays indoor air (top) and thermal mass (bottom) temperature measurements on 1 day, from noon to the next noon, for the window and roof-level vent configuration, where the other configurations show the similar varying trend of the temperatures. 
The measurements highlight the strong diurnal temperature variations, with a stable indoor temperature stratification during the day and a uniform temperature at night. This diurnal variation is governed by the temperature difference between the floor and the roof. The floor exhibits a relatively constant temperature over the entire measurement period, while the walls and the roof exhibit a diurnal variation in the temperature. On the walls, a smaller diurnal variation of up to 7$^o$C occurs. Three of the four walls exhibit very similar behavior, but the south wall, which is adjacent to a shared outdoor kitchen, reaches slightly higher temperatures between noon and 2pm.  The roof experiences large temperature fluctuations up to 20$^o$C due to radiation. During the day, solar radiation heats the roof to set up the stable stratification; at night, radiation to the sky causes the roof temperature to drop below the outdoor air temperature, such that the stratification ceases to exist. 

\subsection{Ventilation rate measurements} \label{ach_measurement}
\subsubsection{Measurement method and post-processing}
The local research team performed ventilation rate measurements in the target house using a tracer concentration decay technique. The technique evaluates the ventilation rate in a space by analyzing how fast the concentration of a tracer gas decays. Particulate matter, generated by burning mosquito coils, was used as the tracer because of its easy accessibility and affordability. After burning the coil, the air in the home was allowed to mix during a 5 minute period. Subsequently, the skylight and/or the windows or vents were opened and the tracer concentration decay was measured at the center of the space. A particulate matter monitor (AM520 SidePak Personal Aerosol Monitor, TSI incorporated) with a sampling frequency of 1 Hz was used. 

The calculation of the instantaneous ventilation rate is based on the relationship that the concentration of a tracer is expected to decay exponentially from its peak value as time elapses:
\begin{equation}
    \dot{V}_{nv}(t) =V_{House}\cdot \frac{\log(c(t)) - \log(c_{peak})}{t - t_{peak}},
    \label{ach_equation}
\end{equation}
where $V_{House}$ is the volume of air inside the house, $t$ is the time and $c(t)$ is the concentration of the tracer at that time, while $c_{peak}$ and $t_{peak}$ indicate the peak concentration and  the time at which the peak occurs. The relationship assumes well-mixed conditions with a spatially uniform tracer concentration, as well as negligible values of the tracer in the outdoor environment. These conditions can be difficult to achieve during field experiments, which introduces some uncertainty in the ventilation rates determined using this technique. To reduce and quantify some of this uncertainty,
the signal is processed by calculating a time series of $\dot{V}_{nv}$ using equation~\ref{ach_equation}, and then computing the mean and the standard deviation of this time series, considering a 5- to 10-minute window with a quasi-steady state signal. This approach has the advantage of eliminating the effect of peaks that were observed at the start and end of some of the measured time series, while also providing a measure of the fluctuations observed during a period of quasi-steady state decay. For the rest of analysis, we present ventilation rate in terms of air change per hour (ACH), the ratio of the ventilation rate ($\dot{V}_{nv}$) to the volume of the house ($V_{House}$), because the quantity has two primary advantages. First, ACH [1/hour] represents how many times the indoor volume is replaced every hour on average, so using ACH is more informational and intuitive than using the ventilation rate with the unit of volume flow rate, [$m^3/s$]. Second, ACH is independent to the volume of a house, thus makes it easier to compare the ventilation status of different households.

\subsubsection{ACH data} \label{subsec:ACHdata}
\begin{figure}[hbt]
    \centering
    \includegraphics[width=1.0\textwidth]{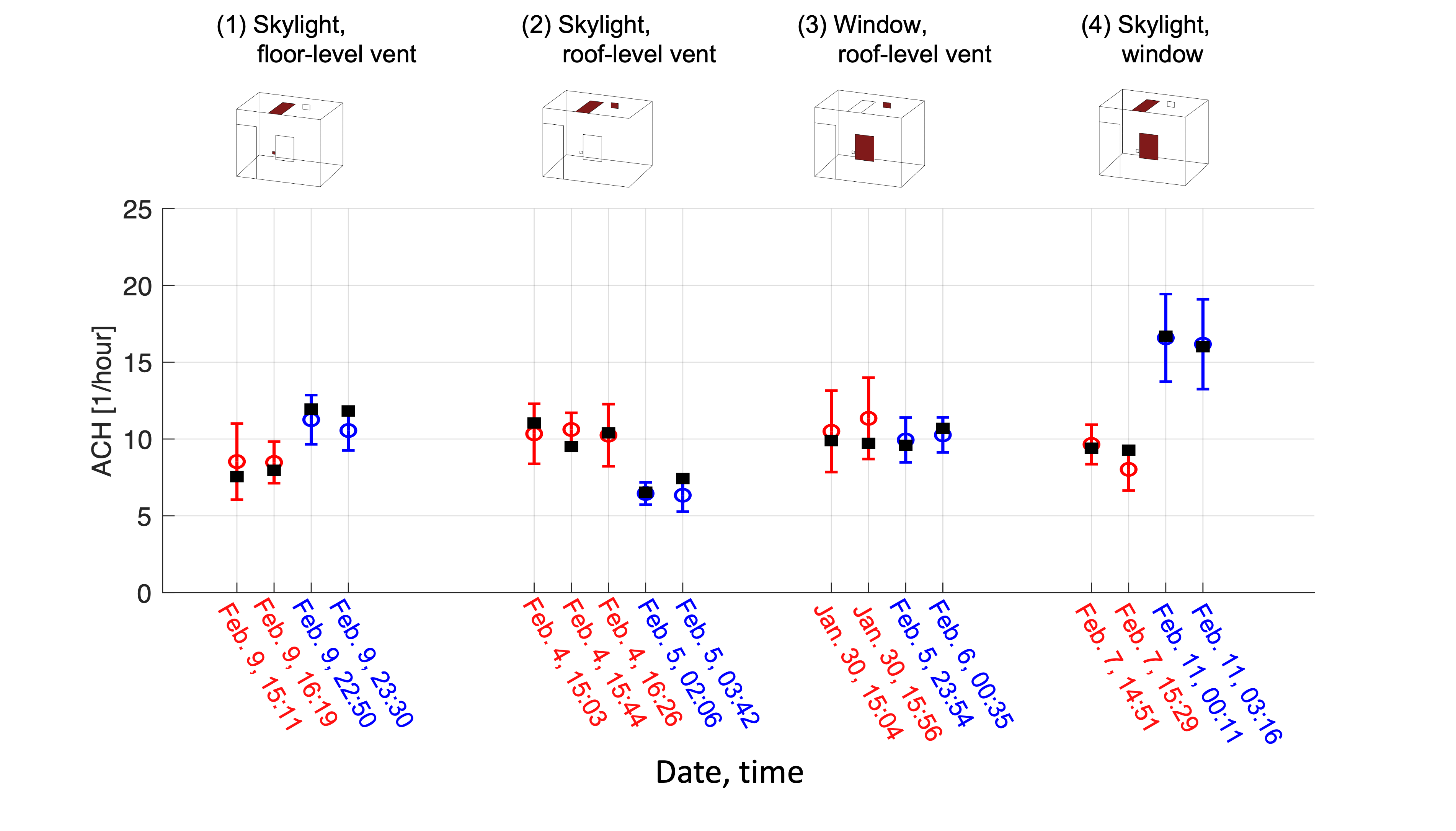}
    \caption{Summary of total 17 ACH measurements using the two post-processing schemes}
    \label{fig:ach_summary}
\end{figure}

Figure~\ref{fig:ach_summary} summarizes the 17 ventilation rate measurement results for the four ventilation configurations used for validation. 
Similar to the temperature measurements, all four configurations have noticeably different ACH values between the daytime and nighttime. These differences can be attributed to two factors: (1) the generally increased wind speeds in the afternoon, and (2) the opposite sign of the indoor/outdoor temperature difference during the day vs the night, together with a day-time stable indoor stratification vs a night-time uniform temperature. When analyzing the relative performance of the different configurations based on the measurements, these differences in the driving forces due to wind and buoyancy at the time of the measurements should be accounted for. 

The exact driving forces that will determine the ventilation rate in a specific configuration depend on many variables, including the wind speed and direction, the indoor/outdoor temperature difference and the surface temperatures that define the boundary conditions for the temperature field in the home. As a starting point, we will assume that the dominant variables in this problem are the wind speed at roof height and the overall indoor to outdoor temperature difference, such that dimensional analysis suggests a functional relationship between the non-dimensional ventilation rate and the ventilation Richardson number $Ri_v$, defined as the ratio of the magnitude of both driving forces:
\begin{equation}
\label{eq:ACHnd}
    \frac{\dot{V}_{nv}}{A_{total}\cdot U_{wind}} = f(Ri_v) = f(\frac{g\cdot\Delta T / T \cdot H}{U_{wind}^2}).
\end{equation}
In this equation, $A_{total}$ is the total area of both openings, and $U_{wind}$ is the wind speed at roof height. The $Ri_v$ numerator represents the buoyant driving force as the product of the gravitational acceleration $g$, the non-dimensional indoor to outdoor temperature difference $\Delta T / T$, and the height of the house $H$. The denominator represents the driving force due to wind as the square of the wind velocity at roof height.

\begin{figure}[htbp]
    \centering
    \includegraphics[width=0.8\textwidth]{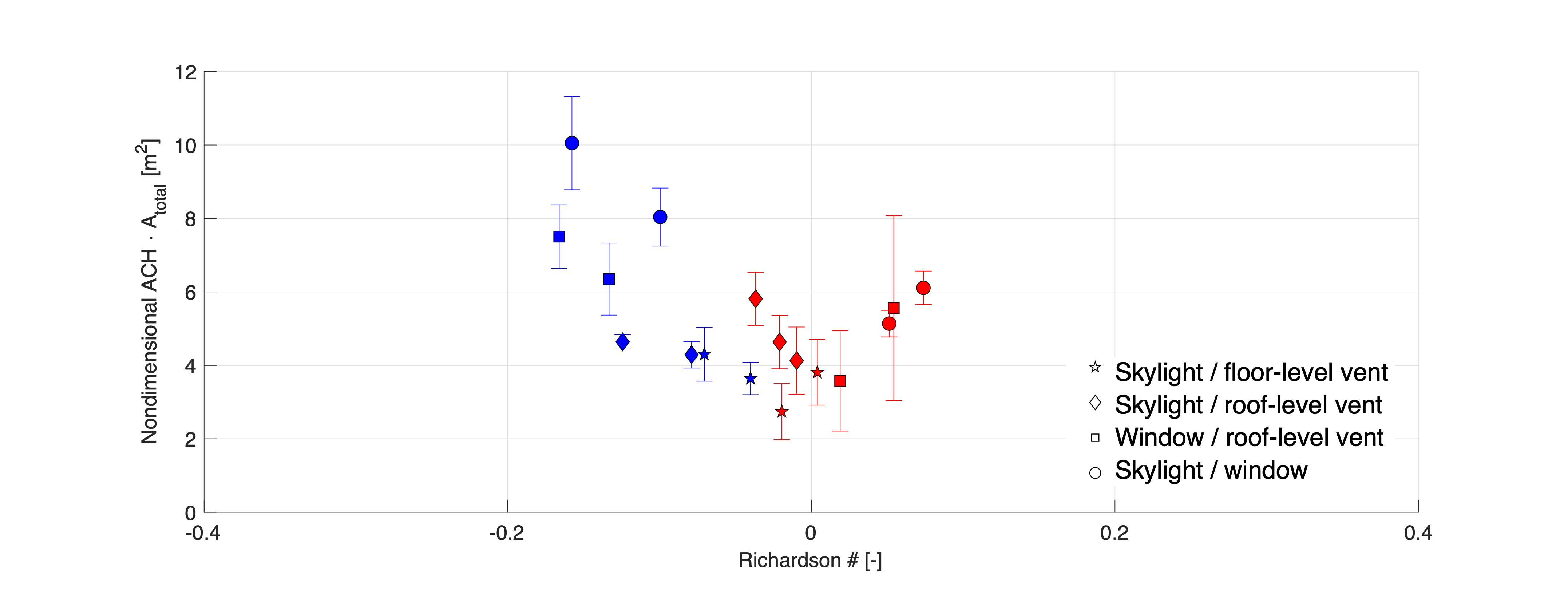}
    \caption{Non-dimensional ACH times total area of openings as a function of ventilation Richardson number, a non-dimensional number that relates the magnitude of two driving forces: wind and buoyancy}
    \label{fig:ach_measurement_area_vs_ri}
\end{figure}

By plotting the non-dimensional ACH multiplied by the total opening area against $Ri_v$, as shown in Figure~\ref{fig:ach_measurement_area_vs_ri}, the performance of the different ventilation configurations can be compared. The four different markers indicate different ventilation configurations, while different colors indicate daytime (red) vs nighttime (blue) measurements. During the day, the outdoor temperature is mostly larger than the indoor air temperature, such that $Ri_v$ is mostly positive. During the night, the outdoor temperature is lower than the indoor temperature, and the Richardson numbers are negative. The plot reveals a strong dependency of the measured values on $Ri_v$ for all configurations; this will be leveraged in Section~\ref{subsec:ventilationmodel} to propose an empirical correlation for the non-dimensional ventilation rate. The limited number of measurements makes it more challenging to identify consistent differences between the different ventilation configurations; the main conclusion from this analysis is that the window and skylight configuration, which has the largest opening area, will provide the highest ACH during the night.  

\section{Description of the building thermal model with uncertainty quantification} \label{computational model}

\begin{figure} [b!]
    \centering
    \includegraphics[width=1.0\textwidth]{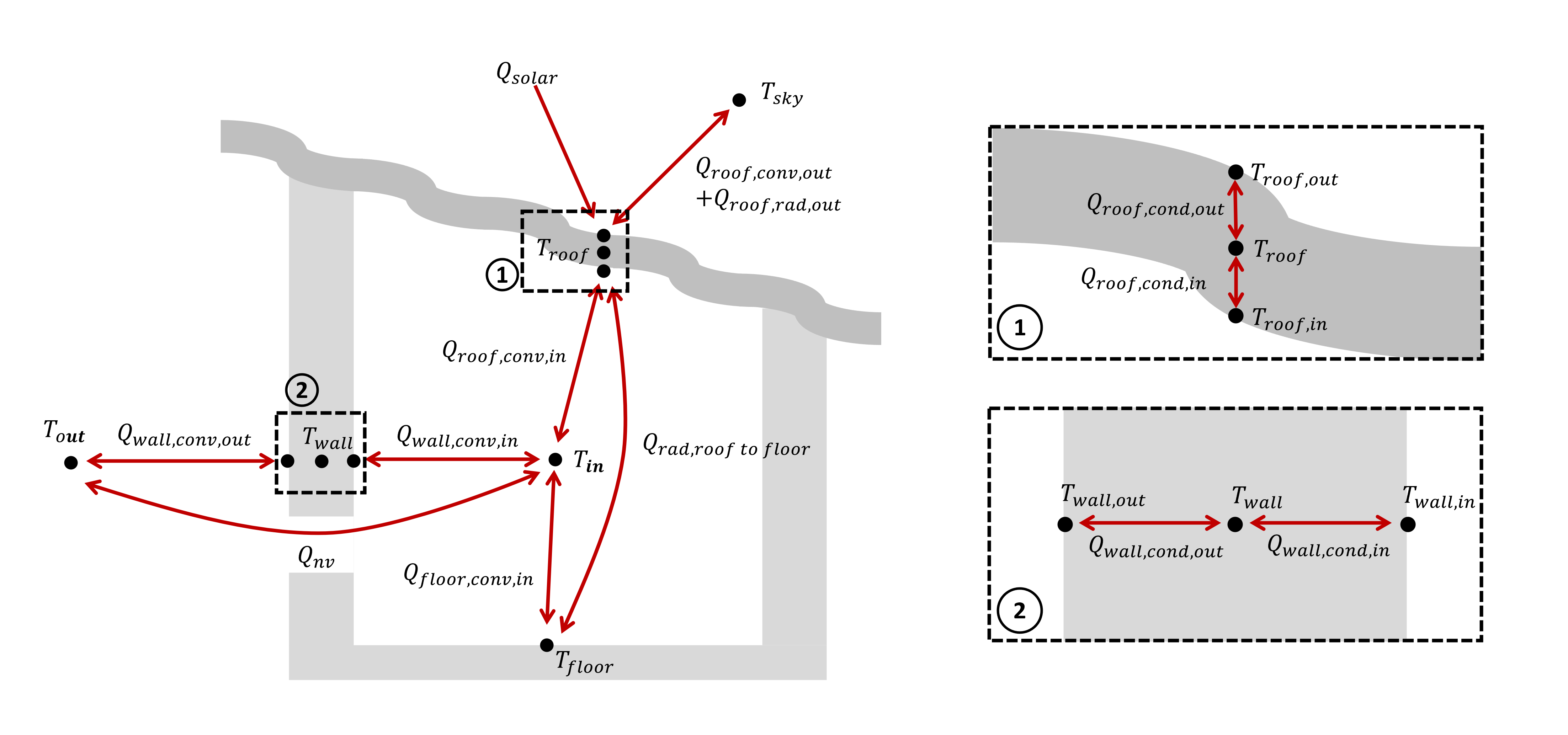}
    \caption{Schematics of governing equations that models heat transfer in the target house: indices in equation \ref{eq:governing_equation} are replaced with actual names of the thermal masses: roof (i=1), wall (i=2,3,4) and floor(i=5).}
    \label{fig:heat_eq_schematics}
\end{figure}
The building thermal model is implemented in OpenModelica, an open-source environment for computational simulations. The governing equations represent the physics that dominate the heat transfer between the indoor air, the roof, floor and walls, and the outdoor environment, as depicted in figure~\ref{fig:heat_eq_schematics}. 
This section first presents the governing equations for the air temperature (section 3.1) and the roof and surface temperatures (section 3.2). Subsequently, the different ventilation models, used to represent epistemic uncertainty in the calculated ventilation rate, are introduced (section 3.3) and the uncertainty due to the variable weather conditions is characterized (section 3.5). Section 3.6 summarizes all uncertain parameters and outlines the propagation method.

\subsection{Governing equations of the building thermal model} \label{governing_equations} 

\subsubsection{Governing equation for the indoor air temperature}
The time evolution of the indoor air temperature $T_{in}$ is governed by heat transfer due to convection $Q_{i,conv,in}$ and natural ventilation $Q_{nv}$:

\begin{equation}
\rho_a V_a C_{p,a} \frac{dT_{in}}{dt} = \sum_{i=1}^{5} Q_{i,conv,in} + Q_{nv},  \label{eq:governing_equation} \\ 
\end{equation}
where $\rho_a$, $V_a$ and $C_{p,a}$ are the density, the volume, and the specific heat of the indoor air, respectively. 

Convective heat exchange is considered for five surfaces ($i=1:5$): the floor, the roof, and three walls. The fourth wall, which is shared with a neighboring home is assumed to be adiabatic due to symmetry. The convective heat transfer with these surfaces is given by:
\begin{equation}
Q_{i,conv,in} = h_{in} A_{i} (T_{i,in}-T_{in}). \\
\end{equation}
$h_{in}$, the convective heat transfer coefficient, depends on the indoor flow pattern and is defined as an uncertain parameter with a uniform distribution between 1 and 4~\citep{lamberti2018uncertainty}. $A_i$ and $T_{i,in}$ are the surface area and temperature, respectively. For the floor, the indoor surface temperature is assumed to be constant. The equations to calculate the roof and wall temperatures are introduced in Sec.~\ref{subsec:thermalmassT}. 

The heat exchange by natural ventilation $Q_{nv}$ is modeled using:
\begin{equation}
Q_{nv} = \rho_a C_{p,a} \dot{V}_{nv} (T_{out}-T_{in}).
\label{eq:Qnv}
\end{equation}
$Q_{nv}$ is proportional to the product of the ventilation flow rate $\dot{V}_{nv}$ and the temperature difference between the outdoor and indoor environments $T_{out}-T_{in}$; additional air exchange by unintended leakage is neglected.
$\dot{V}_{nv}$ is obtained using empirical envelope flow models.To represent the epistemic uncertainty in these models, an ensemble approach that considers different models, introduced in section 3.3 will be employed.

\subsubsection{Governing equations for the roof and wall temperatures}
\label{subsec:thermalmassT} 
 The roof and wall surface temperatures are calculated by solving for the time evolution of their core thermal mass temperature $T_i$, which is governed by conduction through two thermal mass layers, namely $Q_{i,cond,in}$ on the indoor side and $Q_{i,cond,out}$ on the outdoor side:
\begin{equation}
\rho_{i} V_{i} C_{p,i} \frac{dT_{i}}{dt} = Q_{i,cond,in} + Q_{i,cond,out},
\label{eq:goveq2}
\end{equation}
\[\begin{dcases}
Q_{i,cond,in} &= k_{i}A_{i} \frac{T_{i, in} - T_{i}  }{0.5 t_{i}}, \\ Q_{i,cond,out} &= k_{i}A_{i} \frac{T_{i, out}- T_{i}  }{0.5 t_{i}}.  \end{dcases}\]
$\rho_i$, $V_i$, $C_{p,i}$, $t_{i}$ and $k_i$ are the density, the volume, the specific heat, the thickness and the conductivity of the $i$-th thermal mass.

Eqs.~\ref{eq:governing_equation} and~\ref{eq:goveq2} are coupled through the boundary conditions for the heat fluxes at the thermal mass surfaces. On the walls, the conductive heat fluxes into the thermal mass should balance the convective heat fluxes between the wall surfaces and the air; on the roof they should balance the sum of the convective and radiative heat fluxes: 
\begin{align*}
&\text{Wall} \begin{cases}
 Q_{wall,cond,in} &=Q_{wall,conv,in}\\
 Q_{wall,cond,out} &=Q_{wall,conv,out}
\end{cases}  \\
&\text{Roof}  
\begin{cases}
Q_{roof,cond,in} &=Q_{roof,conv,in}+Q_{rad,roof\,to\,floor} \\ 
Q_{roof,cond,out} &=Q_{roof,conv,out}+ Q_{solar} + Q_{rad,sky},
\end{cases}
\end{align*}
The convection terms on the indoor walls and the roof can be calculated following the previously introduced expression for $Q_{i,conv,in}$. The heat convection on the outdoor wall and roof surfaces is modeled equivalently, but using an outdoor heat transfer coefficient $h_{out}$ and the outdoor air temperature $T_{out}$. $h_{out}$ depends on the external flow (wind speed and direction), and is assumed to be uncertain parameter with a uniform distribution between 1 and 15~\citep{emmel2007new}. $T_{out}$ also has an associated uncertainty, introduced in Section 3.4. 

The indoor long-wave radiation term from the roof to the floor is modeled using:
\begin{equation}
\label{eq:rad} Q_{rad,roof\,to\,floor} = \sigma \epsilon_{i} A_{i} (T_{roof,in}^4 -T_{floor}^4)
\end{equation}
with $\sigma$ the Stephan-Boltzmann constant and $\epsilon_i$ the emissivity of the indoor thermal mass surface. Given uncertainty in the roof material properties, $\epsilon_i$ is assumed an uncertain parameter with a uniform distribution between 0.8 and 0.9~\citep{desideri2018handbook}.
The solar radiation acting on the outer roof surface is 
given by:
\begin{equation}
  Q_{solar} = (1-\rho_{roof})A_{roof}I_{solar},
\end{equation}
where $\rho_{roof}$ is the reflectance of the roof and $I_{solar}$ is the solar intensity. Similar to the emissivity, the reflectance is considered an uncertain parameter with a uniform distribution between 0.6 and 0.75~\citep{desideri2018handbook}; the characterization of the uncertainty associated with the solar intensity is introduced in Section~\ref{sec:input_uncertainty}. 
Lastly, radiation to the sky is an important cooling mechanism during the night. It is calculated similarly to Eq.~\ref{eq:rad}, using the sky temperature $T_{sky}$ calculated from Swinbank's equation~\citep{swinbank1963longwave}:
\begin{equation}
  T_{sky} = 0.0553 (T_{out})^{1.5}.
\end{equation}

\subsection{Ventilation modeling approaches}
\label{subsec:ventilationmodel}

The ideal ventilation scenario for the homes is considered to be cross-ventilation, with one opening consistently providing an inlet and the other an outlet. However, in reality, the highly variable wind- and buoyancy-induced pressure forces might result in a range of flow patterns for any given configuration, and this uncertainty should be reflected in the ventilation models. In this light, we investigate two approaches to represent ventilation model uncertainty: an ensemble modeling approach based on existing ventilation models, and a Richardson fit model approach that leverages the ACH measurements performed in the home. 

\subsubsection{Ensemble modeling approach}
\label{subsubsec:ensemblemodel}
The ensemble modeling approach uses two different ventilation models to calculate $\dot{V}_{nv}$ in Eq.~\ref{eq:Qnv}, representing either cross ventilation between two openings or single-sided ventilation through each opening independently. Furthermore, two different scenarios are considered within the cross-ventilation model, namely wind forces assisting vs opposing buoyancy forces. Using this ensemble approach, three BTM solutions are obtained for each set of input parameters. The equations for these ventilation models are introduced in the following paragraphs.

\paragraph{Cross ventilation model}
The ventilation rate of a well-mixed space with two openings $\dot{V}_{nv}$ under cross-ventilation can be calculated following the analytical solution of a steady-state envelope flow model \citep{ventilation_hunt}:
\begin{equation}\label{eq:cross_ventilation}
  \dot{V}_{nv} = A_{eff} \cdot \sqrt{|g\cdot \Delta H \cdot \frac{\Delta T}{\overline{T}}  + U_{wind}^2\frac{\Delta C_p}{2}|}, 
  \end{equation}
  with the effective area $A_{eff}$ calculated from the size $A_{i}$ and the discharge coefficient $C_{d,i}$ of both openings ($i = 1,2$):
 \begin{equation}
  A_{eff} = \frac{A_1 \cdot A_2}{(\frac{1}{2\cdot C_{d,1}^2\cdot C_{d,2}^2}(C_{d,1}^2\cdot A_1^2 + C_{d,2}^2\cdot A_2^2) )^{1/2}}.
\end{equation}
The first term under the square root in Eq.~\ref{eq:cross_ventilation} represents the effect of buoyancy, with $g$ is the gravitational constant, $\Delta H$ the height difference between the two openings, and $\Delta T$ and $\overline{T}$ the difference and the average of the indoor and the outdoor temperatures. The second term represents the effect of wind pressure differences, with $U_{wind}$ the wind speed at the reference height, and $\Delta C_p$ the difference between the pressure coefficients at the two openings. $\Delta C_p$ is a function of the specific geometrical configuration and the wind direction; it can assume a positive (wind and buoyancy assist each other) or negative (wind and buoyancy oppose each other) value. In the present analysis, we ran the model using two different values, namely $\Delta C_p = 0.5$ and $\Delta C_p = -0.5$; these values can reasonably be assumed to provide an upper and lower estimate of the expected cross-ventilation flow rate~\citep{richards2001wind}. More accurate estimates of $\Delta C_p$ could be obtained from a CFD analysis or wind tunnel experiments, which will be considered in future work.

\paragraph{Single-sided ventilation model}
Under certain conditions, the pressure difference between two openings might be too small to consistently drive a cross-ventilation flow. In that case, single-sided ventilation could occur across each opening independently, and the total ventilation rate could be calculated as the sum of the single sided flow rates through each opening:
\begin{equation}
\dot{V}_{nv} = \dot{V}_{single,1} + \dot{V}_{single,2}.
\label{eq:singlecombined}
\end{equation}
In single-sided ventilation the effect of turbulence becomes an important factor and a wide range of empirical models has been proposed to represent this effect \citep{warren1978ventilation, de1982ventilation, dascalaki1996combination, larsen2008single}. In the current study, we adopt the Phaff and De Gids model\citep{de1982ventilation}, which quantifies the effects of buoyancy, wind and turbulence as follows: 
\begin{equation}\label{eq:single_sided_ventilation}
\dot{V}_{single}= \frac{1}{2}\cdot A \cdot \sqrt{C_1 \cdot U_{wind}^2 + C_2 \cdot H \cdot |\Delta T| + C_3}.
\end{equation}
In Eq.~\ref{eq:single_sided_ventilation}, $A$ and $H$ are the area and the height of an opening, $U_{wind}$ is the wind speed at the reference height, and $\Delta T$ is the temperature difference between the indoor and the outdoor environment. The three coefficients ($C_1=0.001 [-]$, $C_2=0.035 [m/s^2 \cdot K]$ and $C_3=0.01 [m^2/s^2]$ ) were obtained by fitting to full-scale experiments, balancing the significance of the three components: wind, buoyancy, and (turbulent) diffusion. This model has been widely used to estimate ventilation rates for single-sided ventilation configuration, and has been shown to produce an average error of 30\% in full-scale experiments~\citep{caciolo2011full, larsen2008single}. However, it is noted that the use of dimensional coefficients, in particular the $C_3$ constant, poses an important limitation to the universality of this empirical model. 

\subsubsection{Richardson fit ventilation model}\label{subsubsec:ri_ventilation_model}
As an alternative to the previously introduced ensemble approach, the ACH measurements introduced in section~\ref{subsec:ACHdata} can be leveraged to develop a model specific to the target house. Following the dimensional analysis introduced Eq.~\ref{eq:ACHnd}, the relationship between the non-dimensional ACH and the ventilation Richardson number $Ri_v$ is plotted in Figure~\ref{fig:ach_measurement_area_vs_ri} for all ventilation configurations. A common trend in the data across all ventilation configurations is observed; this trend indicates that the total opening area, the wind speed and the indoor/outdoor temperature difference are the dominant factors determining the ventilation rate, more so than the specific location and individual sizes of the openings. Using the functional form for the analytical solution for cross-ventilation as a starting point, we propose the following functional form for the model: 
\begin{equation}
    \frac{\dot{V}_{nv}}{A_{total}\cdot U_{wind}} 
    =\sqrt{|c_1\cdot \frac{ g \cdot \Delta T/T \cdot H}{U_{wind}^2} +c_2|}+c_3
    = \sqrt{|c_1 \cdot Ri_v +c_2|}+c_3,
\end{equation}
where $c_1, c_2$ and $c_3$ are non-dimensional coefficients equal to ... , respectively. The physical interpretation of these coefficients is as follows: the driving forces due to wind and buoyancy exactly balance each other when the ventilation Richardson number $Ri_v$ is equal to $-c_2/c_1$, and $c_3$ is the non-dimensional ventilation rate due to turbulent air exchange at this balance point. 
\begin{figure}[htbp]
    \centering
    \includegraphics[width=\textwidth]{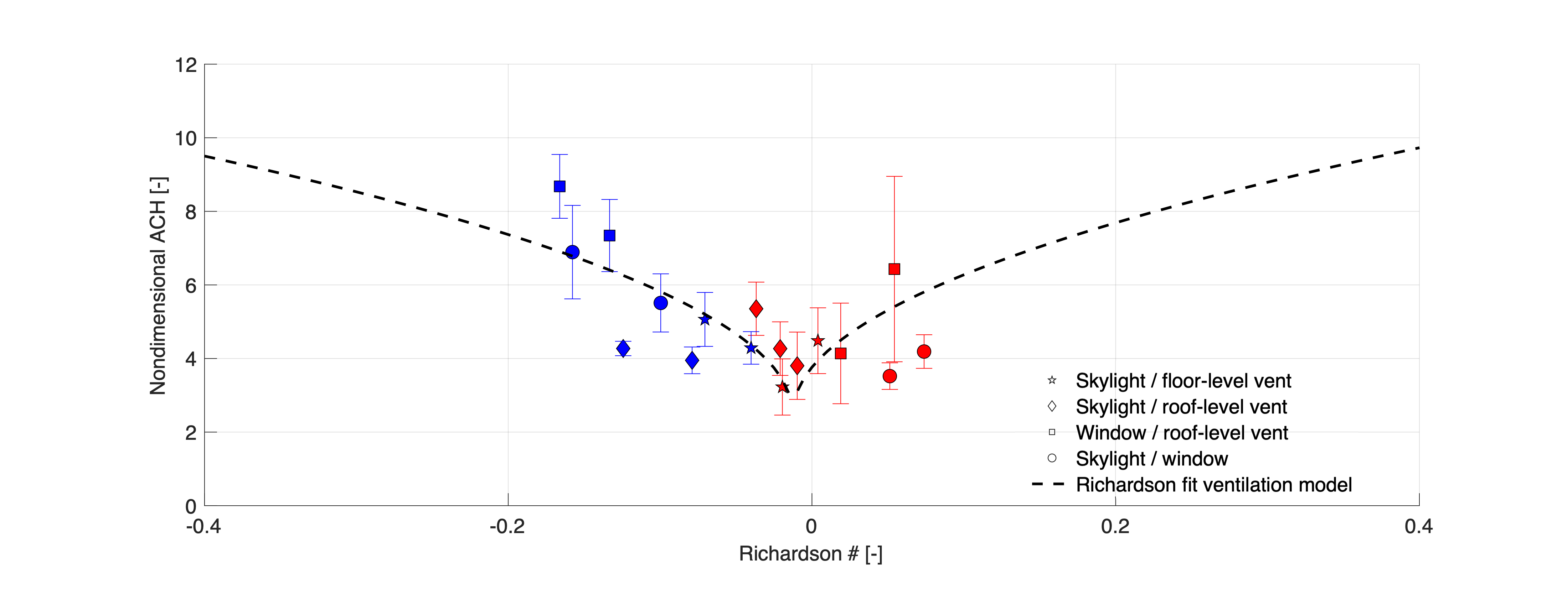}
    \caption{Non-dimensional ventilation rate as a function of Richardson number, and the Richardson fit ventilation model}
    \label{fig:ach_measurement_vs_ri}
\end{figure}

The advantages of fitting this new ventilation model to the measurement data for our specific test case are twofold. First, uncertainty due to a lack of detailed understanding of the ventilation pattern, e.g., cross or single-sided ventilation, is eliminated. Second, the parametric uncertainties related to the pressure coefficient difference in Eq.~\ref{eq:cross_ventilation} or the dimensional coefficients in Eq.~\ref{eq:single_sided_ventilation} can be reduced by obtaining the model coefficients based on the data. As a result, this ventilation model is expected to provide more accurate predictions, even using only a few ventilation rate measurements with different combinations of ventilation openings.

\subsection{Characterization of uncertainty in weather inputs}
\label{sec:input_uncertainty}
\begin{figure}[b!]
    \centering
    \includegraphics[width=1.0\textwidth]{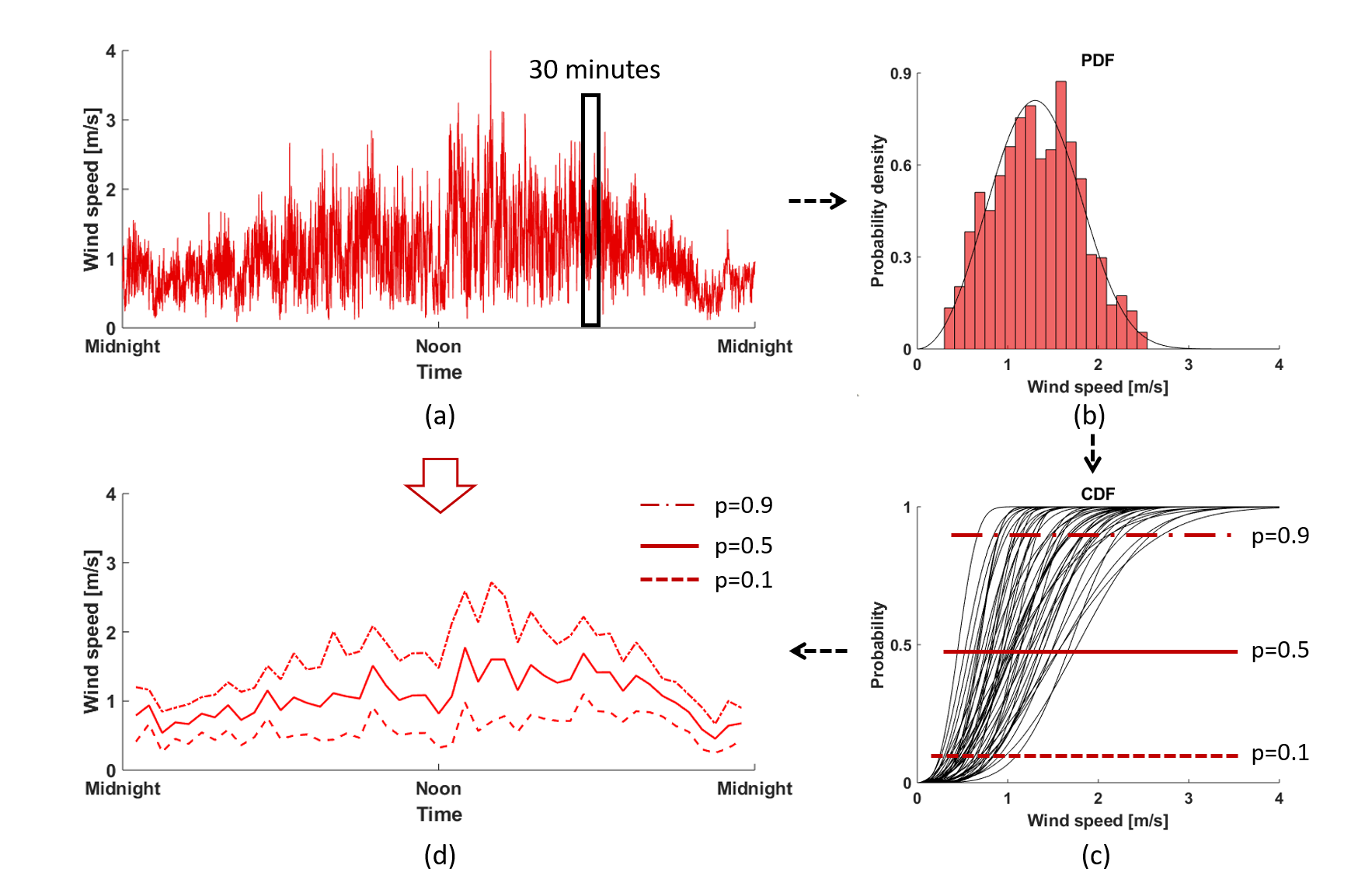}
    \caption{Steps for post-processing the raw weather data to obtain the inputs for the UQ analysis.}
    \label{fig:input_uncertainty}
\end{figure}
The BTM requires three weather inputs in the form of time series of the outdoor temperature, solar radiation, and wind speed at building roof height. Based on the data introduced in Section~\ref{sec:temperature wind measurement}, the temperature and radiation data are obtained from the nearby airport weather station, while the wind speed data is obtained from the anemometer deployed on a local building roof, corrected to the target house roof height. 

The measured time series are first post-processed with a 30-second moving-average filter to average out smaller scale turbulent fluctuations, corresponding to a moderately turbulent wind with a turbulence intensity of 22\%~\citep{bailey1997wind}. Subsequently, we characterize the inherent variability in the weather conditions within 30-minute time frames, corresponding to the duration of the ACH field measurements, as shown in Figure~\ref{fig:input_uncertainty}. This process first converts each 30 minute time series into a probability distribution (Figure~\ref{fig:input_uncertainty}(a) and (b)). This data follows distinct probability density functions for each weather input: (1) the outdoor temperature follows a truncated normal distribution with range $\pm 3.5 \sigma$, determined by its mean and standard deviation; (2) the solar radiation follows a uniform distribution, determined by its minimum and maximum; and (3) the wind speed follows a Weibull distribution, determined by its scale and shape factors. To sample these distributions for the UQ analysis, we use inverse transform sampling (Figure~\ref{fig:input_uncertainty}(c)). 
For each weather input, we draw a sample $p$ from a uniform distribution between 0 and 1, such that the weather input $x$ for that 30 minute period is obtained as $x = F_x^{-1}(u)$, where $F_x^{-1}$ is the inverse cumulative distribution function of $x$. The inputs for the full 24 hours of a BTM simulations are obtained by applying this process for each 30 minute window (Figure~\ref{fig:input_uncertainty}(d)). Each BTM model simulation uses a single value of $p$ for each weather input; this approach will result in conservative bounds on the quantities of interest.

\subsection{Summary of uncertain parameters and propagation method} \label{uncertain_parameters}
\begin{table}[htbp]
\centering
\begin{tabular}[width=\textwidth]{|c| l|c|c}\hline
& \textbf{UQ parameters}  & \textbf{Range} \\ \hline \hline
$h_{indoor}$ & Indoor convective coefficient & [1, 4] \\
$h_{outdoor}$ & Outdoor convective coefficient & [1, 15]  \\ 
$\rho_{roof}$ & Reflectance of the roof &  [0.60, 0.75] \\ 
$\epsilon_{roof}$ & Emissivity of the roof & [0.80, 0.90] \\
$p_{temp}$ & Probability; Indoor temperature & (0, 1) \\
$p_{rad}$ & Probability; Solar radiation & (0, 1) \\
$p_{wind}$ & Probability; Wind speed & (0, 1) \\ \hline 
\end{tabular}
\caption{\label{tab:uq_param} Summary of uncertain parameters for the UQ study}
\end{table}

The uncertainties in the BTM originate from three sources: parametric uncertainty, weather input uncertainty, and ventilation model uncertainty.
Table \ref{tab:uq_param} lists the uncertain parameters related to the first two sources and reports the ranges considered in the UQ analysis. First, there are the four uncertain parameters in the governing equations related to the representation of convective heat transfer and the material properties, as introduced in Section~\ref{governing_equations}. Second, there are three uncertain parameters related to the weather inputs ($p_{temp}$, $p_{rad}$ and $p_{wind}$), as defined in Section~\ref{sec:input_uncertainty}.

For the ventilation model, two different approaches are explored. The first approach, introduced in Section~\ref{subsubsec:ensemblemodel}, quantifies ventilation model uncertainty by running three evaluations of the BTM for each combination of the uncertain input parameters in Table~\ref{tab:uq_param}. These three evaluations use (1) the cross-ventilation model for wind assisting buoyancy, (2) the cross-ventilation model for wind opposing buoyancy, or (3) the single-sided ventilation model. The second approach relies on the Richardson fit ventilation model, which was introduced in Section~\ref{subsubsec:ri_ventilation_model}.

The total number of uncertain parameters to be propagated through the simulations is 7; in addition, this propagation is done for each of the 3 model evaluations in the ensemble modeling approach. Given this large number of uncertain parameters and the low computational cost of a single simulation, simple Monte-Carlo sampling is used to propagate these uncertainties. It was found that 1,000 samples provided converged values for the mean and 95\% CIs of QoIs, i.e. the air and surface temperatures and the ACH. In addition to calculating mean values and CIs, we also conduct a global sensitivity analysis by calculating the first order Sobol indices for the quantity of interest ($Y$) with respect to the $i$-th uncertain parameter ($X_i$): 
\begin{equation}
    S_i = \frac{V_i}{Var(Y)} = \frac{Var_{X_i}(E_{X_{\sim i}}(Y|X_i))}{Var(Y)}.
\end{equation}
In this equation, $E$ and $Var$ indicate the expectation and variance of a variable, and $X_{\sim i}$ indicates the set of all parameters except $X_i$. The Sobol indices range from 0 to 1 and quantify the relative contribution of each variable to the output variance. 

\section{Results}
\label{sec:results}
The presentation of the results is organized by the two quantities of interest (QoI): the temperature and the ventilation rate. Accurate prediction of the indoor temperature is directly relevant to the ACH predictions, since the indoor/outdoor temperature difference determines the importance of buoyant effects in all three ventilation models. For each QoI, we first present the comparison of the computational predictions and the field measurements. Subsequently, we present the sensitivity analysis to identify the dominant uncertain parameters and support further interpretation of the results. 

\subsection{Validation and sensitivity analysis of temperature predictions}\label{result_temperature}

\subsubsection{Validation}\label{result_temperature_validation}
\begin{figure}[htbp]
    \centering
    \includegraphics[width=1.0\textwidth]{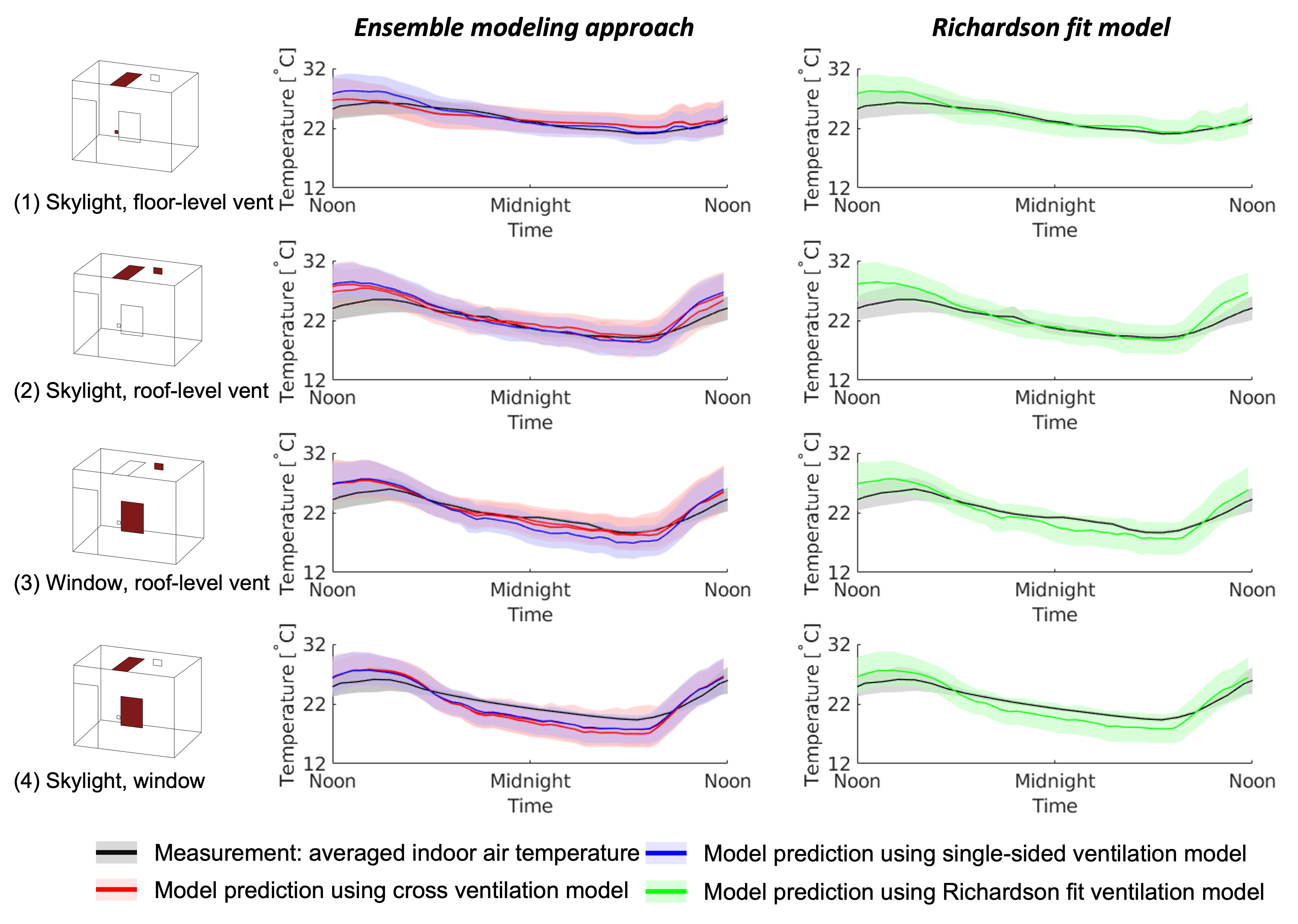}
    \caption{Validation results of the volume-averaged indoor air temperature in the four different ventilation configurations.}
    \label{fig:indoor_temp_validation}
\end{figure}

Figure \ref{fig:indoor_temp_validation} compares the model predictions obtained with the ensemble and Richardson fit ventilation modeling approaches to the measurements in the four different ventilation configurations: (1) skylight and floor-level vent; (2) skylight and roof-level vent; (3) window and roof-level vent; and (4) skylight and window. The measurement data are shown with the black line representing the average of the temperature across all measurement locations; the gray shaded area indicates the 95\% confidence interval (CI), defined as $\pm 1.96 \sigma$, with $\sigma$ the standard deviation of the measurements in the different locations. Similarly, the model predictions are shown with a line representing the mean prediction and a shaded region representing the 95\% confidence interval, as obtained from propagating the uncertainties through each model. The two different predictions using cross ventilation model represent the two cases using $\Delta C_p$ value of 0.5 and -0.5, respectively.


The plots indicate that the air temperature predictions obtained with the different ventilation models are almost identical. The predictions capture the measurement trend throughout the 24h period, generally slightly over-predicting the daytime temperature and slightly under-predicting the nighttime temperature. During the day, the maximum discrepancy between the mean model predictions and the mean of the temperature measurements is 3.4$^\circ$C at 1 pm in the skylight and roof-level vent configuration. During the night, the maximum discrepancy is 2.3$^\circ$C at 3 am in window and roof-level vent configuration.
The CIs of the predictions encompass the spatially averaged temperature throughout the entire day for both models. The magnitude of the CIs is on the order of +/-3$^o$C during the day and +/-2$^o$C at night; the sensitivity analysis presented in Section \ref{result_temperature_sensitivity} will support identifying which uncertain parameters contribute most to the variance in the results.

\begin{figure}[htbp]
    \centering
    \includegraphics[width=1.0\textwidth]{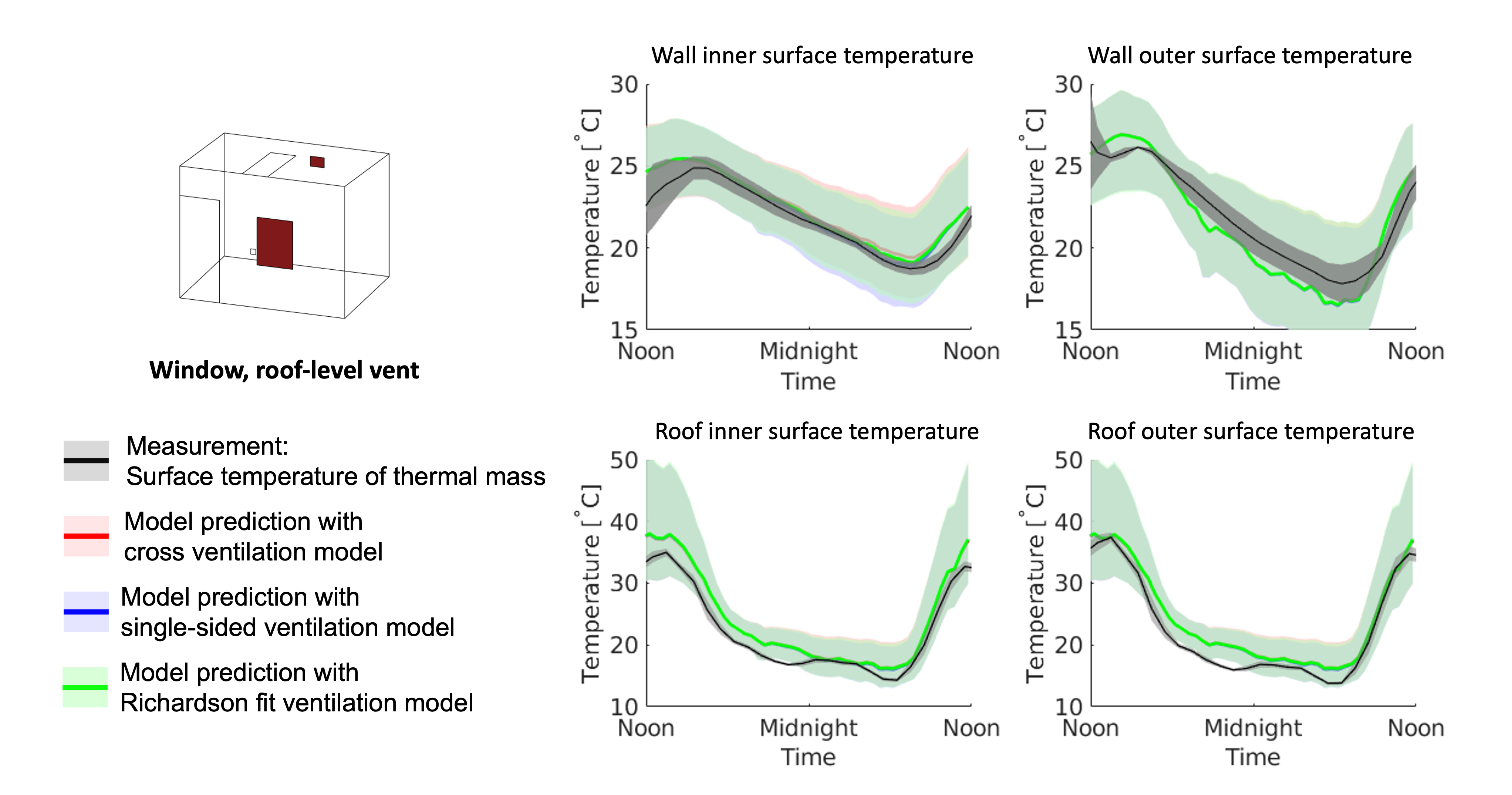}
    \caption{Validation results of thermal mass temperatures in the skylight and roof-level vent configuration: each plot shows the measurements and the three predictions using the three ventilation models.}
    \label{fig:tm_temp_validation}
\end{figure}
Figure \ref{fig:tm_temp_validation} presents the comparison of the thermal mass temperatures obtained from the measurements and all ventilation models for the skylight and roof-level vent configuration; this result is representative of all ventilation configurations considered. The predictions with all ventilation models are shown in a single plot, since they are almost identical. The top row shows the measurements and predictions obtained for the inner and the outer wall surface temperatures. The measurements are averaged over the different walls for which indoor and outdoor measurements are available, since the model predicts a single value for all walls that consider heat exchange with the outdoors. Across all ventilation configurations, the maximum discrepancy between the mean model predictions and the averaged measurements is 3.4$^\circ$C during day and 2.4$^\circ$C at night. The bottom row shows the inner and the outer roof temperatures obtained from the measurements and the predictions using the two ventilation models. For the roof surface temperatures, the maximum discrepancy is 10.6$^\circ$C during the day, and 4.6$^\circ$C during the night in skylight and floor-level vent configuration. The confidence intervals predicted by the models are on the order of +/-5$^o$C for wall surfaces and +/-10$^o$C for roof surfaces, and encompass the measurements 98\% and 71\% of the time for wall surfaces and roof surfaces, respectively. 

In summary, the results indicate that the air and wall surface temperature predictions are not affected by uncertainty in the ventilation model, and that accounting for uncertainty in the model parameters and weather conditions produces CIs that encompass most of the experimental data. The observed discrepancies, as well as the uncertainty in the results, are larger during the day than at night; this is an important observation given that night-time predictions, when children are asleep in the home, are of most interest. 

\subsubsection{Sensitivity analysis} \label{result_temperature_sensitivity}

\begin{figure}[htbp]
\centering
    \includegraphics[width=\textwidth]{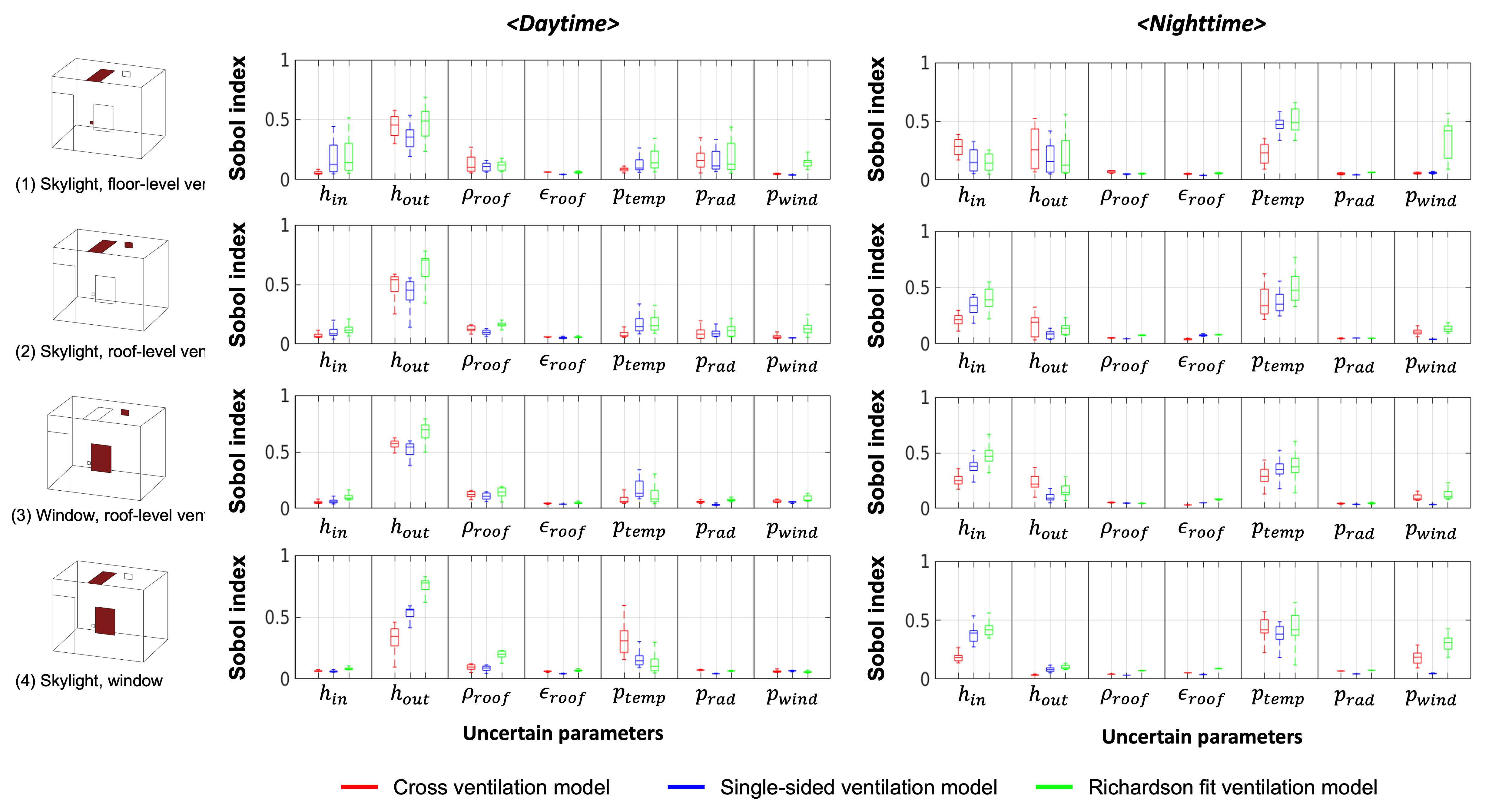}

\caption{Sobol indices of the indoor air temperature in four ventilation configuration(rows) using three ventilation models(colors) during daytime (left) and nighttime(right). The indoor temperature is sensitive to $h_{out}$, $\rho_{roof}$, $p_{temp}$ and $p_{rad}$ during the day; and  $h_{in}$, $h_{out}$ and $p_{temp}$ at night for all configurations and ventilation models used.}
\label{fig:sobol_index_temp}
\end{figure}

Variance-based sensitivity analysis quantifies the relative contribution of the different uncertain parameters to the variance in the prediction of the indoor air temperature. Given the diurnal temperature variation observed in the measurements and the simulations, the analysis is performed separately for the stratified conditions during the day (9 am - 3 pm) and the well-mixed conditions during the night (9 pm - 3 am). During these time windows, Sobol indices are calculated at each 1 min interval. Figure~\ref{fig:sobol_index_temp} presents the distribution of the resulting Sobol indices using 8 plots, where the four rows represent the four ventilation configurations and the two columns represent the daytime and nighttime results. Each plot presents 3 x 7 box plots, color-coded for the 3 different ventilation models, for each of 7 uncertain parameters. For the cross ventilation model, the results using $\Delta C_p$ of 0.5 are presented; the use of $\Delta C_p=-0.5$ leads to similar results. For each box plot, the box indicates 25\% and 75\% percentiles, the lines extending from the box indicate the minimum and the maximum values, and the center line in the box indicates the median (50\% percentile) of the distribution of each parameter, respectively. 

 Considering the daytime analysis, the effect of the external convective heat transfer coefficient ($h_{outdoor}$) is the most pronounced for all ventilation configurations, implying that the outdoor convection is a dominant heat transfer mechanism for determining the indoor air temperature during the day. In addition, depending on the ventilation configuration and the model used, the indoor convective heat transfer coefficient ($h_{in}$), the density of the roof material $\rho_{roof}$, as well as the parameters that define the outdoor temperature and solar radiation $p_{temp}$ and $p_{rad}$, can have non-negligible influences.

For the nighttime analysis, the importance of $h_{out}$ is reduced, while the importance of $h_{in}$ is increased. $p_{temp}$ remains an important factor, and the parameter that provides input related to the wind speed to the ventilation model $p_{wind}$ plays a more important role, in particular in the cross-ventilation and Richardson fit models. The increased importance of the latter parameter implies that the ventilation heat transfer has a more pronounced impact on the indoor temperature during the night than during the day. The only exception to this observation is the skylight and floor-level vent configuration when using the cross-ventilation model; however, as will be discussed in section~\ref{result_ventilation_validation}, this simulation strongly under-predicts the ventilation rate. 

In summary, the sensitivity analysis suggests that reducing uncertainty in the daytime predictions will require a more accurate characterization of the radiative, convective, and conductive heat transfer at the roof, while reducing uncertainty in the nighttime predictions will require a more accurate characterization of the indoor and outdoor convective heat transfer. In addition, the variance in the nighttime predictions can partially be attributed to inherent variability in the wind and temperature conditions that define the natural ventilation flow rate and heat exchange.

\clearpage
\subsection{Ventilation rate: validation and sensitivity analysis} \label{result_ventilation}

\subsubsection{Validation} \label{result_ventilation_validation}
\begin{figure} [h!]
\centering
    \includegraphics[width=\textwidth]{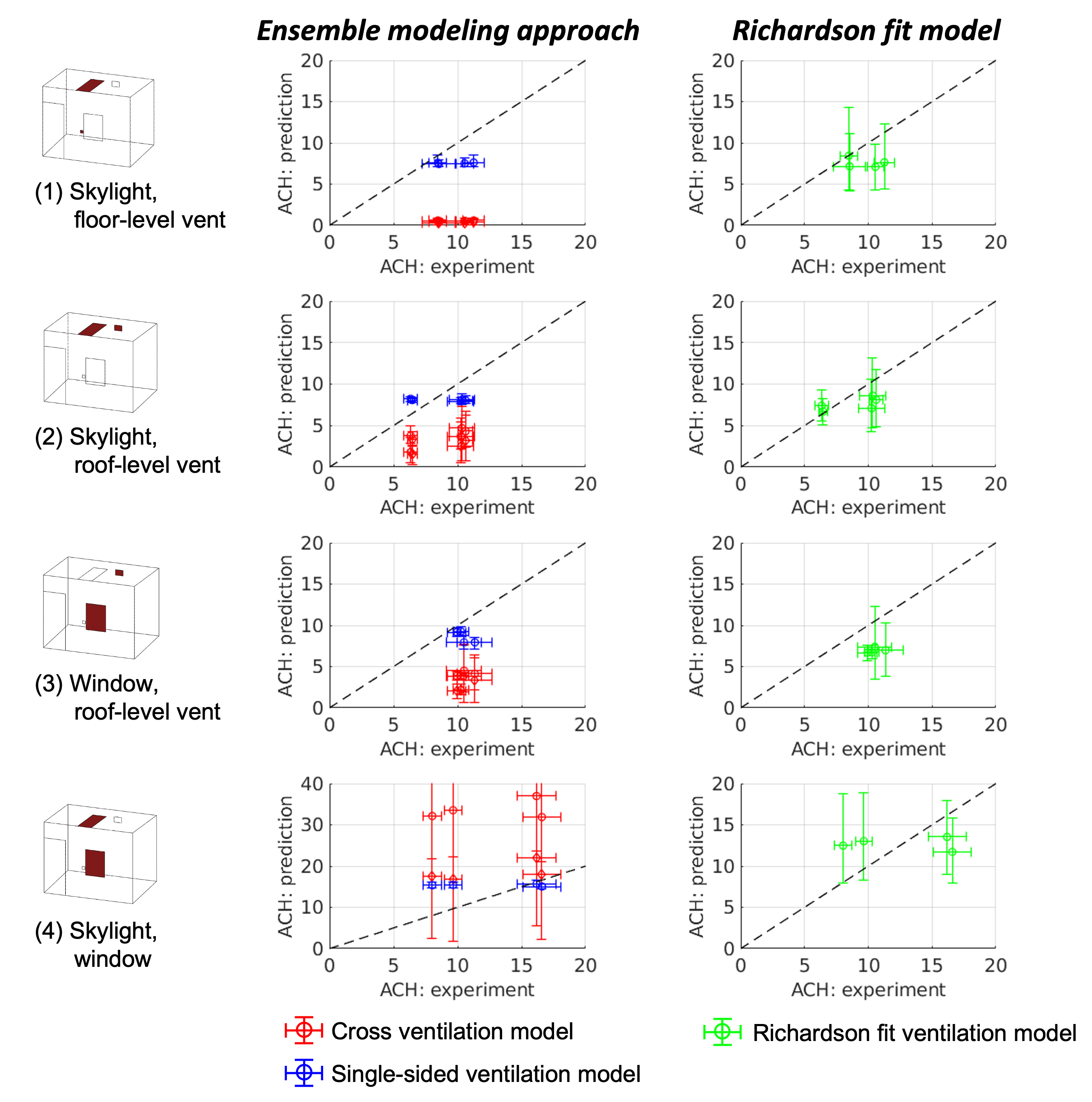}
\caption{Scatter plots of ACH predictions and measurements using the two ventilation modeling approaches}
\label{fig:ach_validation}
\end{figure}

Figure \ref{fig:ach_validation} shows scatter plots comparing the measurements and model predictions in the four ventilation configurations, using the ensemble and Richardson fit modeling approaches. In the scatter plots, the horizontal error bar indicates the 95\% CI of the measurement, calculated using the procedure outlined in section~\ref{ach_measurement}; the vertical error bar indicates the 95\% CI as predicted by the building thermal model with UQ. The model simulations were performed for each of the days during which an ACH measurement was conducted, and the ACH values shown were extracted during the time period of the measurements. The plot indicates significant uncertainty in the predicted ACH for each configuration, both due to uncertainty in the form of the ventilation model and due to uncertainty in the model parameters and weather inputs. The discussion in this section focuses on the differences between the results obtained with the different ventilation models; the sensitivity analysis in section~\ref{result_ventilation_sensitivity} will support identifying which uncertain model parameters and weather inputs contribute most to the variance in each model's predictions. 

The accuracy of predictions using ensemble modeling approach is limited for all configurations. Considering the first three rows, i.e. the skylight and floor-level vent, skylight and roof-level vent, and window and roof-level vent configurations, the ACH predictions are generally lower than the measurements. Despite the large combined confidence intervals for the predictions, ranging from 0 to 9.8 ACH, the confidence intervals only encompass the data in 3 out of 13 measurements performed in these configurations. For the skylight-window configuration the mean model predictions are generally higher than the measurements, but the extremely large confidence intervals, ranging from 2.5 to 55.2 ACH, render the ensemble model prediction non-informative. 

When inspecting the cross-ventilation and single-sided ventilation models individually, some interesting observations about their respective shortcomings can be made. In the configurations with one small and one large opening, the ACH prediction obtained from the cross-ventilation model is limited by the cross-sectional area of the smaller opening. In reality, single-sided ventilation is likely to occur at the large opening. This is reflected in the more realistic ACH magnitudes predicted by the single-sided ventilation model. However, the single-sided model does not correctly represent the effect of varying indoor/outdoor temperature or wind conditions: it predicts very similar values for measurements under different conditions, and the CIs are too narrow to encompass the measured data points. In the configuration with two large openings, the cross-ventilation result is strongly impacted by the assumption made regarding wind assisting or opposing buoyancy. Comparison to the experimental data indicates that in reality, it is unlikely that a steady uni-directional flow is established between both openings. As for the other configurations, the single-sided model performs slightly better than the cross-ventilation model, but its predictions do not reflect the impact of varying indoor/outdoor temperature and wind conditions. Overall, the RMS errors across all experiments are: (1) Cross ventilation model with $\Delta C_p=+0.5$: 12.0; (2) Cross ventilation model with $\Delta C_p=-0.5$: 7.6, and (3) Single-sided ventilation model: 3.08.

 In contrast to the ensemble modeling approach, the confidence intervals predicted with the Richardson fit model encompass the measurement data for 12 out of 17 measurements. In the remaining 5 measurements the difference is small (1.5 ACH on average). The RMS error across all experiments is similar to the single-sided ventilation model (3.04), but the more realistic confidence intervals indicate that the model has an improved capability to capture changes in ACH due to varying weather conditions. The results reveal a general under-prediction of the higher ACH measurements. This discrepancy could be reduced by obtaining more data to support an improved fit, potentially considering different fits for each ventilation configuration and introducing uncertainty in the fit coefficients that can be propagated through the model. 

\subsubsection{Sensitivity analysis} \label{result_ventilation_sensitivity}
\begin{figure}[hbt]
\centering
    \includegraphics[width=0.9\textwidth]{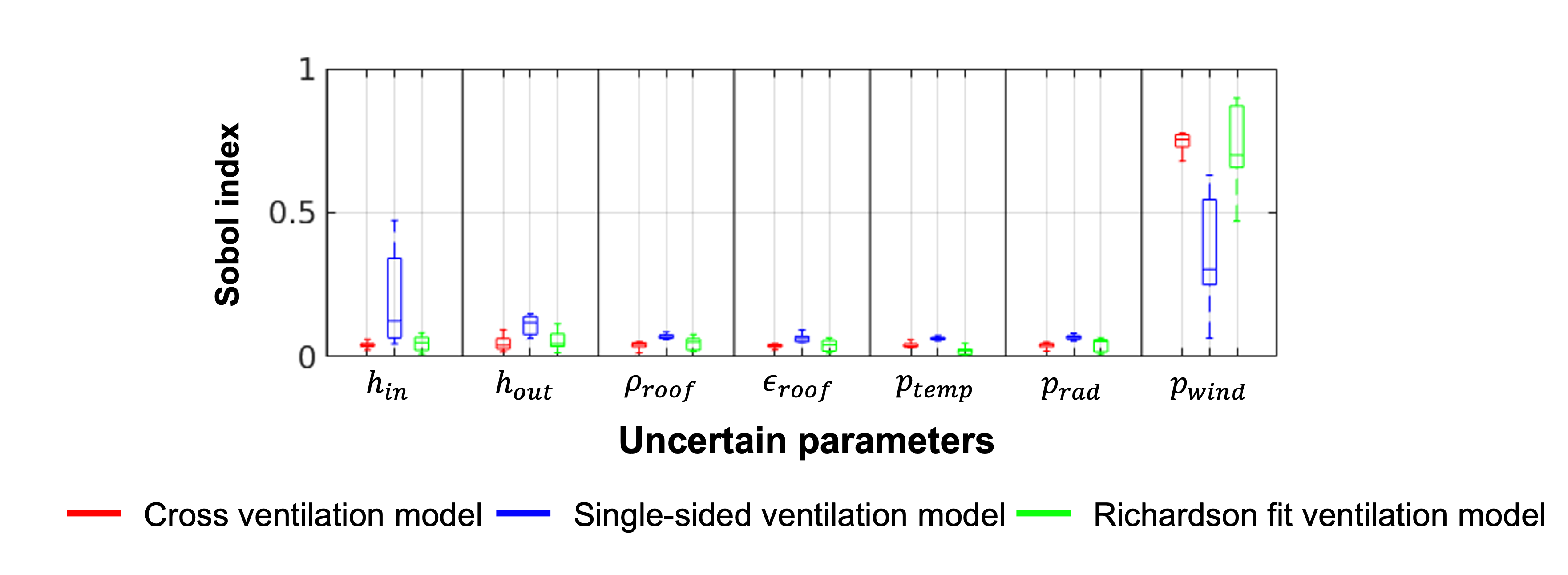}
\caption{Sobol indices of uncertain parameters}
\label{fig:sobol_index_ach}
\end{figure}
Similar to the sensitivity analysis for the temperature predictions in section \ref{result_temperature_sensitivity}, a sensitivity analysis is performed for the ventilation rates predicted by each ventilation model. Figure~\ref{fig:sobol_index_ach} displays the box plots for the Sobol indices obtained across all 17 predictions. The plot indicates that most of the variance in the results obtained with a single ventilation model can be attributed to uncertainty in $p_{wind}$, which is directly associated to the wind-related driving force. For the single-sided ventilation model, a somewhat increased influence of the indoor and outdoor heat transfer coefficient is observed, but when interpreting these results it is important to keep in mind that the overall variance in the ACH predictions from the single-sided ventilation is low. 

In summary, the variance in the predictions obtained with each ventilation model is primarily related to the inherent variance in the wind speed during the 30-minute time periods over which the ventilation rates are calculated. Since this is an irreducible uncertainty, the primary route to improving the ACH predictions us to reduce the epistemic uncertainty in ventilation model by improving our understanding of the natural ventilation process as a function of the weather conditions. This understanding could be obtained from additional field measurements, or from more detailed computational fluid dynamics solutions of the natural ventilation flow.

\section{Conclusion and future work}
This paper has presented a computational framework for predicting indoor temperatures and ventilation rates in slum homes. The framework uses a building thermal model in which the ventilation rate is calculated using envelope flow models that take the indoor to outdoor temperature difference and wind speed as inputs. Furthermore, the framework incorporates uncertainty quantification to account for uncertainty in model parameters, weather inputs, and the form ventilation model. Importantly, the framework was designed to be computationally efficient such that it can quantify ventilation rates in a variety of homes under a variety of weather conditions to support further investigations of the association between the occurrence of pneumonia and household ventilation. 

Validation with on-site field measurements for different ventilation configurations in a representative home indicates that the thermal model correctly captures the daily variation in the average indoor air temperature. The maximum observed discrepancies between the mean predictions and the averaged measurements is 3.4$^\circ$C during the day and 2.4$^\circ$C at night. By accounting for uncertainty in model parameters and weather inputs the model predicts confidence intervals for the indoor air temperature that encompass most of the field measurement data; in contrast, the choice of ventilation model form has a negligible impact of the temperature predictions. This observation changes drastically when considering the ventilation rate expressed in air changes per hour (ACH); as expected, the ACH is strongly affected by the choice of the ventilation model form. Standard cross-ventilation or single-sided ventilation models fail to reproduce the measured values or the trend in the measured values, respectively. When combining the predictions with both models into an ensemble modeling approach the confidence intervals only encompass the measurement data for 7 out of 13 data points. Furthermore, for 4 of these data points the confidence intervals are on the order of +/-20 ACH, such that the results are no longer informative. As an alternative, the available measurement data was used to propose a site-specific ventilation model. This model expresses the dependency of the non-dimensional ventilation rate on the ventilation Richardson number, which quantifies the ratio of the driving forces due to buoyancy and wind. Using this model, the predicted confidence intervals encompass the measurement data for 12 out of 17 measurements. In the remaining 5 measurements the difference is small at 1.5 ACH on average. 

In summary, the results indicate the promising predictive capabilities of a building thermal model with uncertainty quantification, provided that some information on the ventilation rates at the specific site is available. In the present paper, the use of only 17 ACH measurements, obtained across 4 different ventilation configurations, supported suggesting an empirical correlation for a ventilation model that provides ACH predictions with representative confidence intervals. It is expected that the predictive capability of the model can be further improved by obtaining a better understanding of the natural ventilation process as a function of the weather conditions. In future work, we will explore the use of CFD to gain this insight and develop more accurate ventilation models that can be implemented in building thermal models.

\section*{Acknowledgement}
This research was funded by a seed grant from the Stanford Woods Institute Environmental Venture Projects program and supported by the Stanford Center at the Incheon Global Campus (SCIGC) funded by the Ministry of Trade, Industry, and Energy of the Republic of Korea and managed by the Incheon Free Economic Zone Authority.

\clearpage
\bibliographystyle{apalike2}\biboptions{authoryear}
\bibliography{reference.bib}

\end{document}